\documentclass[12pt,referee]{aastex}
\shorttitle{Nanoflare variability with Hinode/XRT}
\shortauthors{Terzo et al.}

\begin{document}

\title{Widespread nanoflare variability detected with Hinode/XRT in a solar active region}

\author{Sergio Terzo\altaffilmark{1}$^\ast$, Fabio Reale\altaffilmark{1}, Marco Miceli\altaffilmark{1}}
% \author{Fabio Reale\altaffilmark{1}}
% \author{Marco Miceli\altaffilmark{1}}
\affil{Dipartimento di Fisica, Universit\`a di
       Palermo, Sezione di Astronomia, Piazza del Parlamento 1, 90134 Palermo,
       Italy}
% \email{terzo@astropa.unipa.it}
\author{James A. Klimchuk}
\affil{NASA Goddard Space Flight Center, Greenbelt, MD 20771, USA}
\author{Ryouhei Kano, Saku Tsuneta}
% \author{Saku Tsuneta}
\affil{National Astronomical Observatory, Mitaka, Tokyo 181-8588, Japan}
\altaffiltext{1}{INAF - Osservatorio Astronomico di Palermo ``G.S.
       Vaiana'', Piazza del Parlamento 1, 90134 Palermo, Italy,
\\$^\ast$E-mail: terzo@astropa.unipa.it}
% normalsize{$^\ast$E-mail: terzo@astropa.unipa.it}

\begin{abstract}
It is generally agreed that small impulsive energy bursts called
nanoflares are responsible for at least some of the Sun's hot
corona, but whether they are the explanation for most of the
multi-million degree plasma has been a matter of ongoing debate.  We
here present evidence that nanoflares are widespread in an active
region observed by the X-Ray Telescope on-board the Hinode mission.
The distributions of intensity fluctuations have small but important
asymmetries, whether taken from individual pixels, multi-pixel
subregions, or the entire active region. Negative fluctuations
(corresponding to reduced intensity) are greater in number but
weaker in amplitude, so that the median fluctuation is negative
compared to a mean of zero. Using MonteCarlo simulations, we show
that only part of this asymmetry can be explained by Poisson photon
statistics. The remainder is explainable with a tendency for
exponentially decreasing intensity, such as would be expected from a
cooling plasma produced from a nanoflare.
We suggest that
nanoflares are a universal heating process within active regions.
\end{abstract}


\section{Introduction}

How the outer atmosphere of the Sun, the solar corona, is heated to
several million degrees Kelvin is one of the most compelling
questions in space science \citep{Klimchuk_2006}.  Simple thermal
conduction from below is clearly not the answer, since the corona is
more than two orders of magnitude hotter than the solar surface.
Indeed, whatever mechanism heats the corona must do so in the face
of strong energy {\it losses} from both downward thermal conduction
and radiation.

Soft X-ray and EUV images of the corona reveal many beautiful loop
structures---arched magnetic flux tubes filled with plasma. It is generally agreed that warm loops -- whose temperature is {\it only} about 1 MK, well observed in EUV images -- are bundles of unresolved thin strands that are heated by small
energy bursts called nanoflares (\citealp{parker_1988,Gomez_93ApJ,Warren_2002ApJ,Klimchuk_2006,Sakamoto_2008}).
Identifiable warm loops account for only a small fraction of the coronal plasma, however.  Most emission has a diffuse appearance, and the question remains as to
how this dominant component is heated, especially in the hotter
central parts of active regions.  Is it also energized by
nanoflares, or is the heating more steady?  Recent observations have
revealed that small amounts of extremely hot plasma are widespread
in active regions \citep{Reale_2009ApJ} and are consistent with the
predictions of theoretical nanoflare models
\citep{Klimchuk_2008ApJ}.  This suggests that nanoflare heating may
indeed be universal.  However, the conclusion is far from certain
\citep{Brooks_2009ApJ}.  The work reported here sheds new light on
this fundamental question.

A magnetic strand that is heated by a nanoflare evolves in a well
defined manner.  Its light curve (intensity vs. time) has a
characteristic shape:
the intensity rises
quickly as the nanoflare occurs, levels off temporarily, then enters
a longer period of exponential decay as the plasma cools (\citealp{L_Fuentes_2010ApJ}).
If we could isolate
individual strands in real observations, it would be easy to
establish whether the heating is impulsive or steady. Unfortunately,
this is not the case.  The corona is optically thin, so each line of
sight represents an integration through a large number overlapping
translucent strands.  Nonetheless, it may be possible to infer the
presence of nanoflares.

Actual light curves exhibit both long and short-term temporal
variations.  Some of the short-term fluctuation is due to photon
statistical noise, but some may be caused by nanoflares.  The
amplitude of the fluctuations seems to be larger than expected from
noise alone \citep{Sakamoto_2008,Sakamoto_2009ApJ,Vekstein_2009A&A}.
However, this is difficult to determine with confidence,
because the precise level of noise depends on the temperature of the
plasma, and this is known only approximately in these studies.  As
we report here for the first time, there is another method for
detecting nanoflares from intensity fluctuations that does not
depend sensitively on the noise.
If heating is impulsive we expect the light curves of individual
strands to be asymmetric.
The strand is bright for less time than it is
faint, and when it is bright it is much brighter than the temporal
average. This results in a distribution of intensities that is also
very asymmetric.  A good measure of the asymmetry is the difference
between the median and mean values.
This is a generic property of light curves that are dominated by an
exponential decay, as is the case with nanoflares. We use this
property to demonstrate that nanoflares are occurring throughout a
particular active region that we studied in detail. Since the light
curve at each pixel in the image set is a composite of many light
curves from along the line-of-sight, the asymmetries of the
intensity distributions and the differences between the median and
mean values are small.  We use both statistical analysis and
quantitative modeling to show that the differences are nonetheless
significant and consistent with widespread nanoflaring in the active
region.

In Section~\ref{sec:data} we describe the data analysis and results, in Section~\ref{sec:model} we interpret the results in the light of Monte Carlo simulations and of loop modeling and in Section~\ref{sec:disc} the whole scenario is discussed.

\section{Data analysis}
\label{sec:data}

\subsection{The observation and preliminary analysis}

The grazing-incidence X-Ray Telescope (XRT)
\citep{Golub_2007,Kano_2008SoPh,Narukage_2011SoPh} on the Hinode
spacecraft \citep{Kosugi_2007} detects plasmas in the temperature
range $6.1 < \log~T<7.5$ with 1 arcsec spatial resolution. Active
region AR 10923 was observed on 14 November 2006 near the center of
the solar disk. It was also studied previously in other ways
\citep{Reale_2007Sci,Reale_2009ApJ}. The observations used for this
study were made in the Al\_poly filterband starting at 11~UT and
lasting $\sim 26$ min.  A total of 303 images were taken with a 0.26
s exposure at cadence intervals between 3 and 9 s. No major flare
activity or significant change in the morphology occurred during
this time.  We concentrated on a 256$\times$256 arcsec$^2$ field of
view and used the standard XRT software to calibrate the data.
The images were co-aligned using the jitter information provided with the data.

\subsection{Data cleaning}

Because we are interested in low level systematic variations that
could be indicative of nanoflares, we removed pixels from the
dataset that show phenomena which may obscure the effect we are
attempting to study.
Our analysis is best applied to light curves that are approximately
constant or that exhibit only a slow linear trend. We therefore
excluded pixels that have a low signal or that show macroscopic
variations that might be attributed to cosmic ray hits, microflares
or other transient brightenings, or to slow variations due, for
instance, to local loop drifts or motions.  We discuss each of these
possibilities in turn.

Since we expect the fluctuations that result from episodic heating
to be erratic and of very small amplitude, they may be very
difficult to distinguish from the noise, so we removed all pixels
with an average count rate below $30 DN/s$. This is essentially the
entire dark area outside of the active region proper. These pixels
amount to $\sim 11$\% of the total.
We removed all pixels affected by bright spikes due to cosmic rays
or point-like brightenings. These pixels were identified by the
condition that the signal is at least $1.5$ times the spatial median
of the immediately surrounding pixels \citep{Sakamoto_2009ApJ}. They
represent $\sim 15 \%$ of the total. We also excluded continuous
macroscopic events, i.e. large scale events such as
\emph{microflares}. To this aim, we performed a linear fit of the
pixel light curve and removed the pixels whose intensity became
or exceeded 1.5 times of the bestfit line at any time during the observation. 
These account for $\sim 10 \%$ of the total. Finally, we removed slow 
intensity variations due to displacement or drift of coronal structures along the line of sight.
We used a method based on counting the number of crossings of the bestfit line by the
light curve. If the fluctuations of \emph{m} data points around the
linear fit are completely random, the time profile has \emph{m-1}
possibilities to cross the linear fit, with $0.5$ probability. The
\emph{``number of crossings''} follows a \emph{Binomial}
distribution with a mean of ${(m-1)}/{2}$ and a standard deviation
of ${\sqrt{(m-1)}}/{2}$. Assuming that the duration of intrinsic
intensity fluctuations is shorter than the observing time ($\sim 26$
min), and the duration of the fluctuations due to loop drifts or
motions is comparable with observing time, the number of crossings
due to loop motions should be smaller than $\sqrt{(m-1)}/{2}$. We
removed all pixels where the number of crossings is smaller than the
mean of the binomial distribution ($\sim 7 \%$). At the end of the
cleaning we are left with about $56\%$ of the total number of pixels
as shown in Figure \ref{fig1}.

\subsection{Temporal analysis}
\begin{figure}
\epsscale{.80}
\plotone{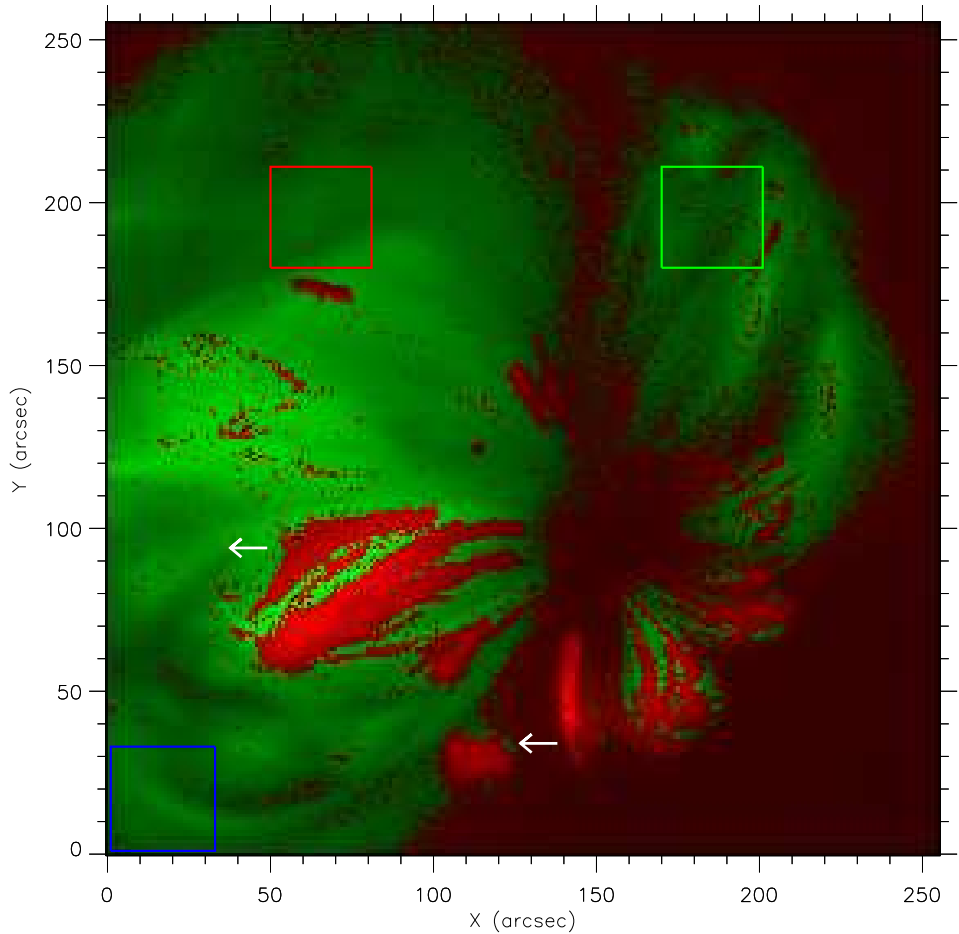}
\caption{Active region AR 10923 observed with the Hinode/XRT Al\_poly filter on 14 November 2006 at 11~UT. We distinguish between pixels accepted (green) and rejected (red) for the analysis. The color scales are powers of the intensity (0.5 and 0.1 for green and red respectively), with maxima of 57 DN/s and 1171 DN/s respectively. We mark three subregions (frames) which are analyzed specifically. We show in Figure \ref{fig2} the light curves of two pixels (indicated by the arrows).}
\label{fig1}
\end{figure}

\begin{figure}
\epsscale{.90} \plotone{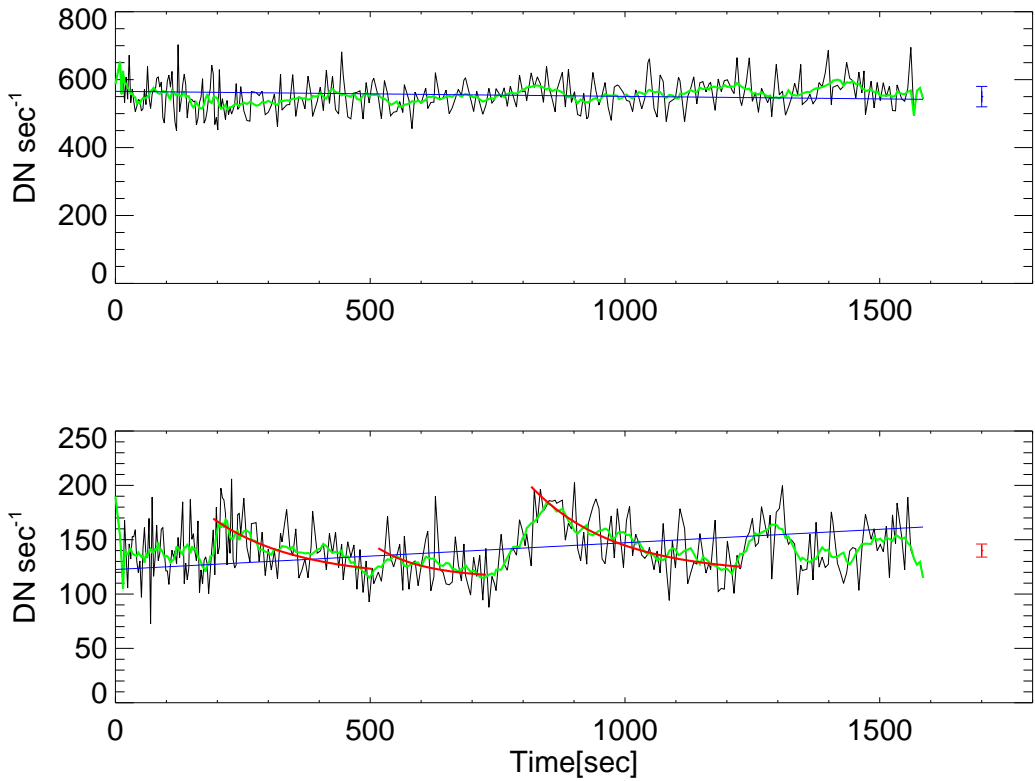} \caption{Light curves of two
selected pixels indicated in Figure \ref{fig1}.  Linear fits are
shown in blue; 9-point ($\sim 1$ min) running averages are shown in
green; in the lower panel we show sample decaying exponentials (red) that fit well some data segments.} \label{fig2}
\end{figure}

\begin{figure}
\epsscale{1.15}
\plottwo{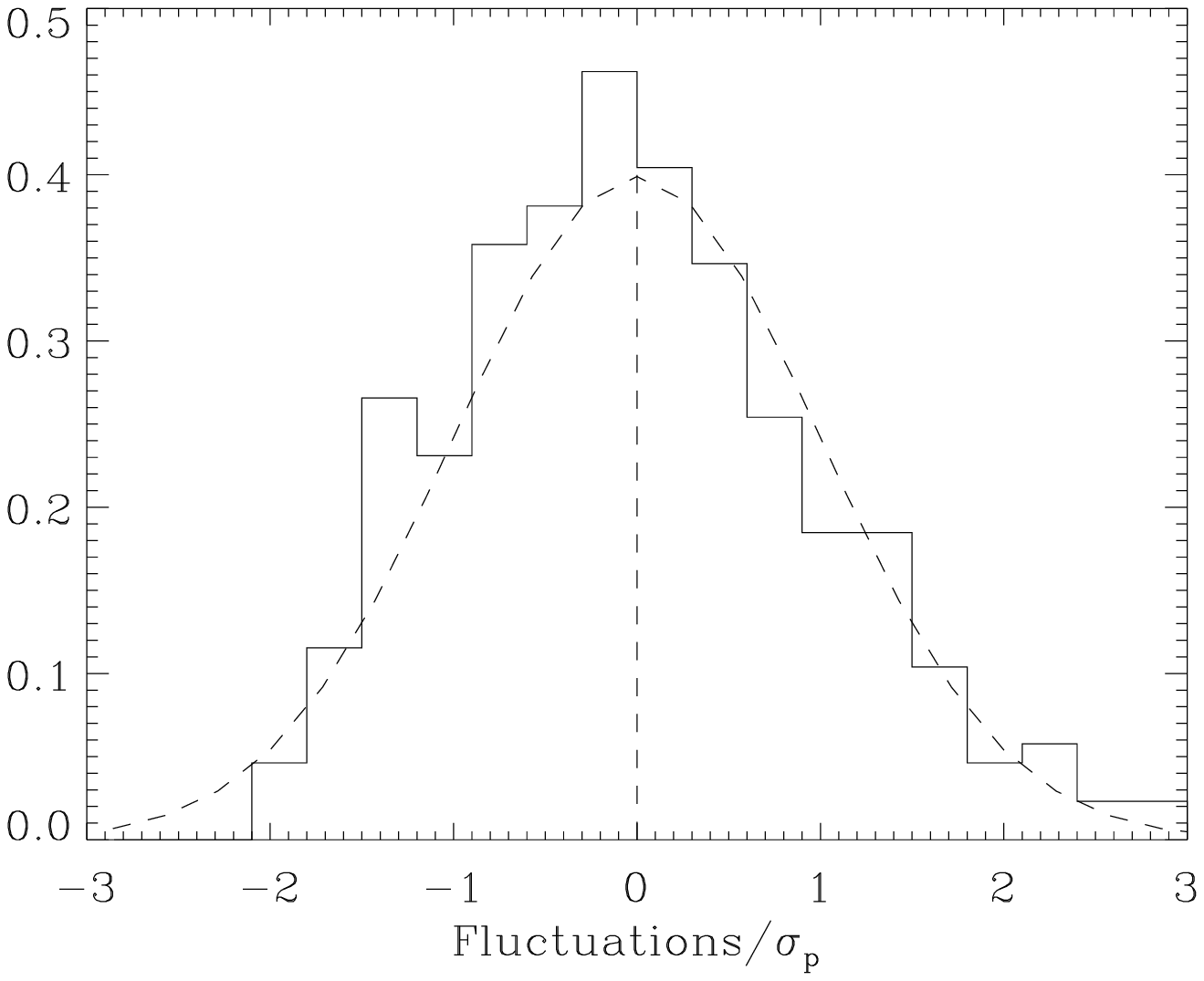}{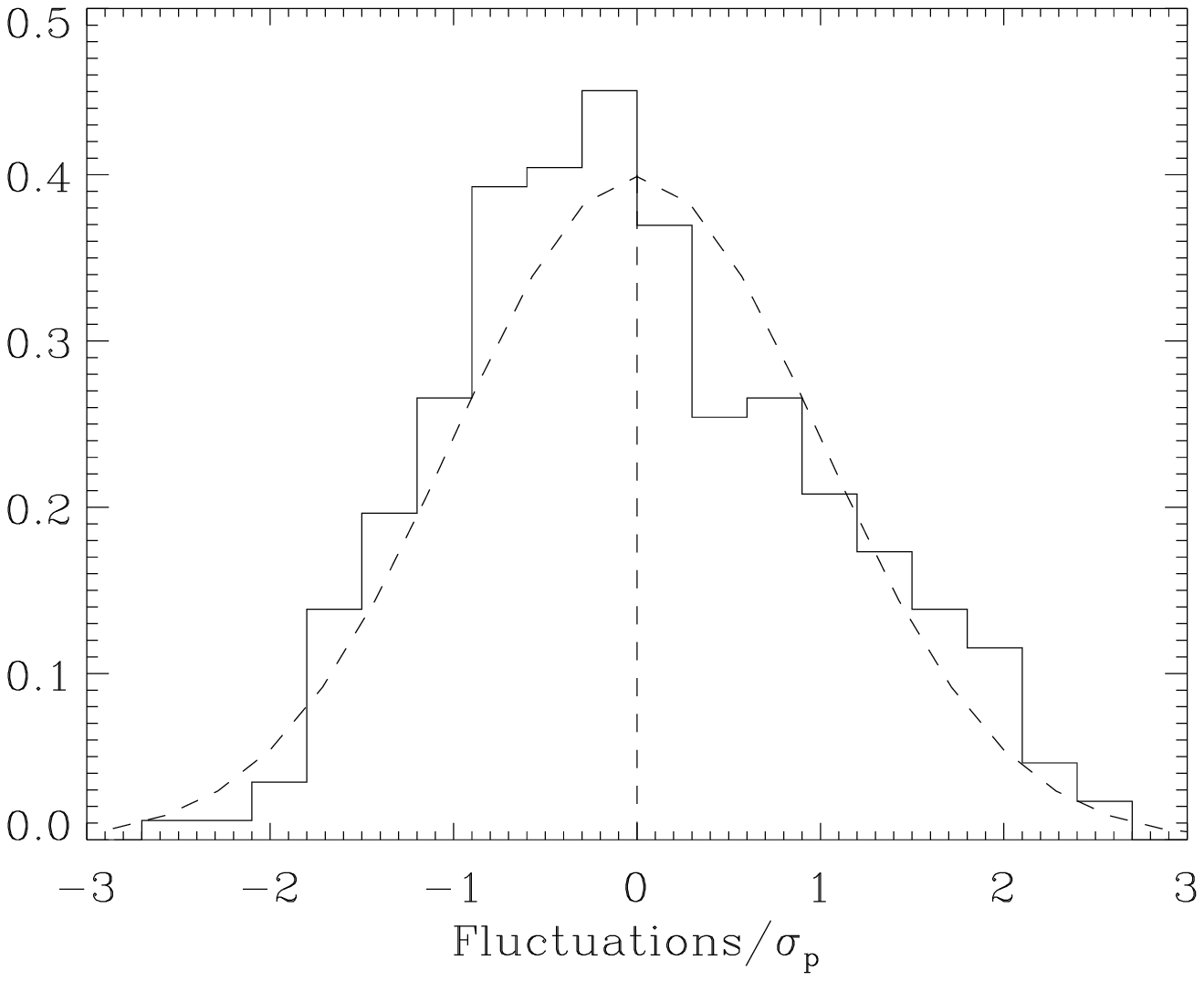}
\caption[]{Distributions of the fluctuations of the light curves with respect to the linear fit in the two selected pixels of Figure \ref{fig1} and \ref{fig2}. The fluctuations amplitude distributions are normalized to the Poisson noise. A Gaussian centered on zero and having unit width is plotted for reference (dashed line).}
\label{fig3}
\end{figure}

The light curves of the remaining pixels (green in Fig.\ref{fig1}) can be fit
satisfactorily well with a linear regression.  The slopes tend to be
very small ($0\pm0.15$ in $90\%$ of the cases), and there is no
preference for increasing or decreasing intensity.  Figure
\ref{fig2} shows light curves for two sample pixels with the linear
fit in blue and 9-point ($\sim 1$ min) running averages in green.
The light curve in the lower panel is one with a highly negative median, 
and on it we mark three decaying exponentials that fit well the respective
data segments and provide good evidence for cooling (see
Sections~\ref{sec:mc},\ref{sec:disc}). We measure intensity
fluctuations relative to the linear fit according to:

\begin{equation}
dI(x,y,t) = \frac{I(x,y,t) - I_0(x,y,t)}{\sigma_P(x,y,t)}
\label{eq:fluc}
\end{equation}
\noindent where $I(x,y,t)$ is the count rate (DN/s) at position
$[x,y]$ and time $t$, $I_0(x,y,t)$ is the value of the linear fit at
the same position and time, and $\sigma_P(x,y,t)$ is the photon
noise estimated as the standard deviation of the pixel light curve
with respect to the linear fit, with a small correction to account for the variation of the average count rate with time (described by the linear fit)\footnote{An alternative possibility is to estimate the photon noise from the nominal relations with signal intensity. These relations require the conversion from DN to photon counts, and therefore depend on the source emitted spectrum. This introduces a strong dependence on the temperature of the emitting plasma. So, to estimate the photon noise in this way one has to make an assumption on the plasma temperature. This is not straightforward in an inhomogeneous active region, and we preferred a model-independent approach.}.
The distribution of the intensity fluctuations (Fig.\ref{fig3}) is not symmetric at either pixel. There is a slight
excess of negative fluctuations (fainter than average emission)
compared to positive.  The mean fluctuation is 0, by definition, but
the median fluctuation (normalized to $\sigma_P$) is $-0.08 \pm
0.07$ in the brighter pixel (upper panel of Fig.~\ref{fig2}) and $-0.12 \pm 0.07$ in the fainter pixel (lower panel of Fig.~\ref{fig2}).
The uncertainties in the median values have been rigorously
computed according to \cite{Hong_2004ApJ}.

Since the fluctuations of each pixel light curve are normalized, in the same way we can build a distribution with higher statistical significance simply including the fluctuations from more pixels.
Figure \ref{fig4} (left
panel) shows the distributions of the three $32 \times 32$ pixels sub-regions marked in Figure
\ref{fig1} and of the whole active region. Subtle asymmetries can
be detected by eye when compared to the Gaussian distribution shown
as a dashed curve for comparison.  The right panel in Figure
\ref{fig4} shows the distributions of the median values themselves,
computed individually at each pixel.  There is a clear preference
for the medians to be negative. The median averages (coinciding with the peak of the median distributions, that are highly symmetric)
are between $-0.025 \pm 0.002$ and  $-0.030 \pm 0.002$ for the
sub-regions and $-0.0258 \pm 0.0004$ for the entire active region. Uncertainties are estimated according to \cite{Hong_2004ApJ}.
Results for the active region and the selected subregions are listed
in Table \ref{tbl-1}.
The fact that the results are similar in the subregions and in the
whole active region (and the significance increases) is important because it shows that the effect is
widespread and real. Were it due simply to random Gaussian
fluctuations (or fluctuations of any random variable that is
symmetrically distributed), the magnitude would decrease as more and
more pixels are included in the statistics, i.e., the effect would
be smaller for the whole active region.
Furthermore, if the effect were due entirely to photon noise, which obeys Poisson statistics (see next Section), then increasing the sample size would bring the Poisson distribution closer to a symmetric Gaussian and decrease the difference between the median and the mean (i.e., bring the mean closer to zero).  However, the measured median is just as large for the entire active region as it is for the sub-regions.

\section{Modeling and interpretation}
\label{sec:model}

\subsection{MonteCarlo simulations}
\label{sec:mc}

Photon counting obeys Poisson statistics, and since the Poisson
distribution is asymmetric, part of the negative offset of the
median values is due to photon noise.  We determine how much by
performing MonteCarlo simulations to generate synthetic light curves
for an appropriate number of pixels.

As a null-hypothesis, we assume that the fluctuations at each pixel
are due only to photon noise, i.e., that the intrinsic light curve
is flat.  To simulate this, we start from an observed emission map
obtained by time averaging all the actual images.  We then introduce
synthetic noise at each pixel using Poisson statistics and having
the same average fluctuation amplitude as observed, derived
according to Equation \ref{eq:fluc}.  In this way we obtain a noisy
light curve, with fluctuations Poisson-distributed around the
zero-value. We repeat this procedure for all valid pixels, thereby
obtaining a datacube of artificial XRT images exactly analogous to
the real one. We can then apply the same analysis to the synthetic
data. As already mentioned, we obtain asymmetric distributions from
the null-hypothesis. For the three subregions marked in Figure
\ref{fig1} we obtain median values between $-0.013 \pm 0.002$ and
$-0.018 \pm 0.002$. These values are incompatible with and
significantly lower than those measured from the observational data
($-0.025/-0.030 \pm 0.002$). For the whole region we obtain $-0.0164
\pm 0.0004$ to be compared to $-0.0258 \pm 0.0004$ from the data.
Analogously we have computed that for all pixels with an average
rate $\geq 800$ and $\geq 1600$ DN/s the median distribution for the
whole region is $-0.0096 \pm 0.0009$ and $-0.0096 \pm 0.0017$,
respectively, to be compared with observational data ($-0.0160 \pm
0.0009$ and $-0.0136 \pm 0.0018$) for the same threshold values
respectively.

Our next step is to perturb the intrinsically flat light curves with
a sequence of random segments of exponential decays, linked one to
the other. We slightly reduce the constant offset so as to maintain
the same average DN rate after adding the perturbations, which are
all positive. The parameters of the perturbations are the
$e$-folding time, $\tau$, the average time interval between two
successive perturbations, $dt$, and the amplitude, $A$. The
$e$-folding time is fixed for each simulation. The cadence is
Poisson-distributed around the average value, because each perturbation is triggered  an integer number of frames after the previous one. Since the number of frames is relatively large (tens) the Poisson distribution approaches a Gaussian one. The amplitude is random-uniform between 0.5 and 1.5 of the average value. 

The flat light
curve becomes ``\emph{saw-toothed}'', but non-periodic, with
exponential descending trends. This new light curve is then
randomized according to the pixel average counting statistics, as
was done for the constant light curve (Fig.\ref{fig5}). Again, we
repeat this procedure for all valid pixels to obtain new datacubes,
which we analyze as if they were real data. 

We perform a sample
exploration of the parameter space. In particular, we consider
reasonable loop cooling timescales as possible $e$-folding times,
i.e. $\tau = 180,\;360,\;540\;s$. The larger values more likely for
realistic active region loops of length $5 - 10 \times 10^{9}$ cm,
according to the loop cooling times ($\tau_{s}$), which are of the
order of \citep{Serio_1991A&A}:

\begin{equation}
\tau_{s} = 4.8 \times 10^{-4} \frac{L}{\sqrt{T_0}}
= 120 \frac{L_9}{\sqrt{T_{0,7}}}
\label{eq:serio}
\end{equation}

where $L$ ($L_9$) is the loop half-length (in units of $10^9$ cm)
and $T_0$ ($T_{0,7}$) is the loop maximum temperature (in units of
$10^7$ K). To give a significantly negative median, each exponential must be visible uninterrupted for a relatively long time, even more since its amplitude is relatively small with respect to the constant background. Therefore we have set the average time interval between two successive perturbations to a value compatible with the chosen $e$-folding time.
We make two different sets of simulations with amplitude A = $30$ and $60\;DN/s$.

\begin{figure}
\epsscale{1.} \plotone{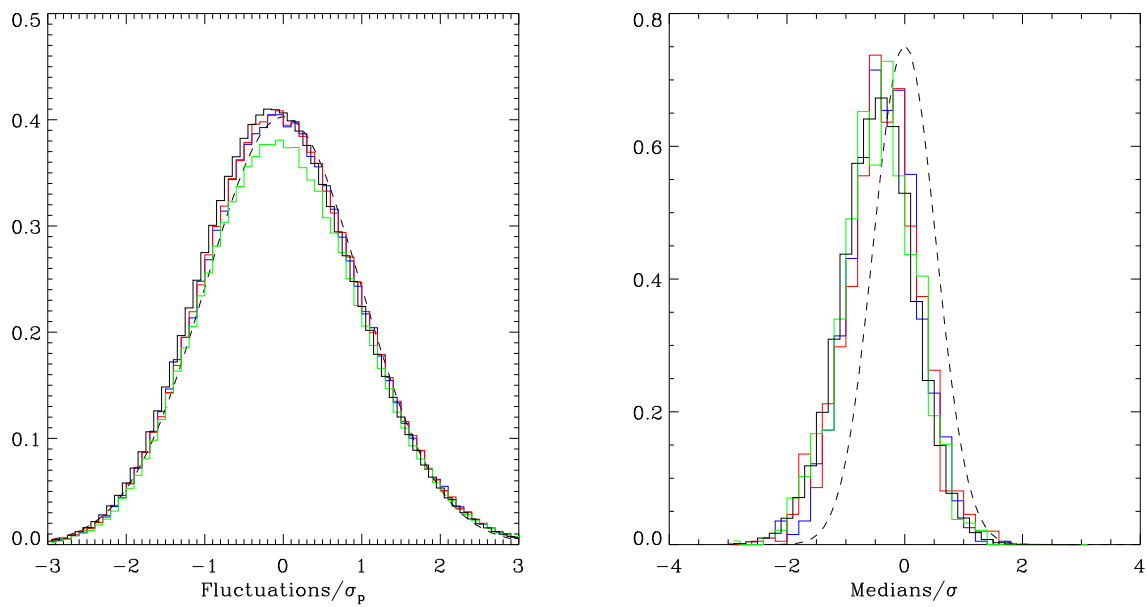} \caption{The left panel shows the
combined distributions of fluctuations for the pixels in three
selected regions (color coded to match the boxes in Figure
\ref{fig1}) and in the whole active region (black histogram). The right
panel shows the distributions of the median fluctuation values
computed individually at each pixel. Fluctuations are normalized to
the Poisson noise (left), and medians are normalized to
their standard deviation (right). Gaussians centered on zero and
having unit width are plotted for reference (dashed line).}
\label{fig4}
\end{figure}

\begin{figure}
\epsscale{0.9}
\plotone{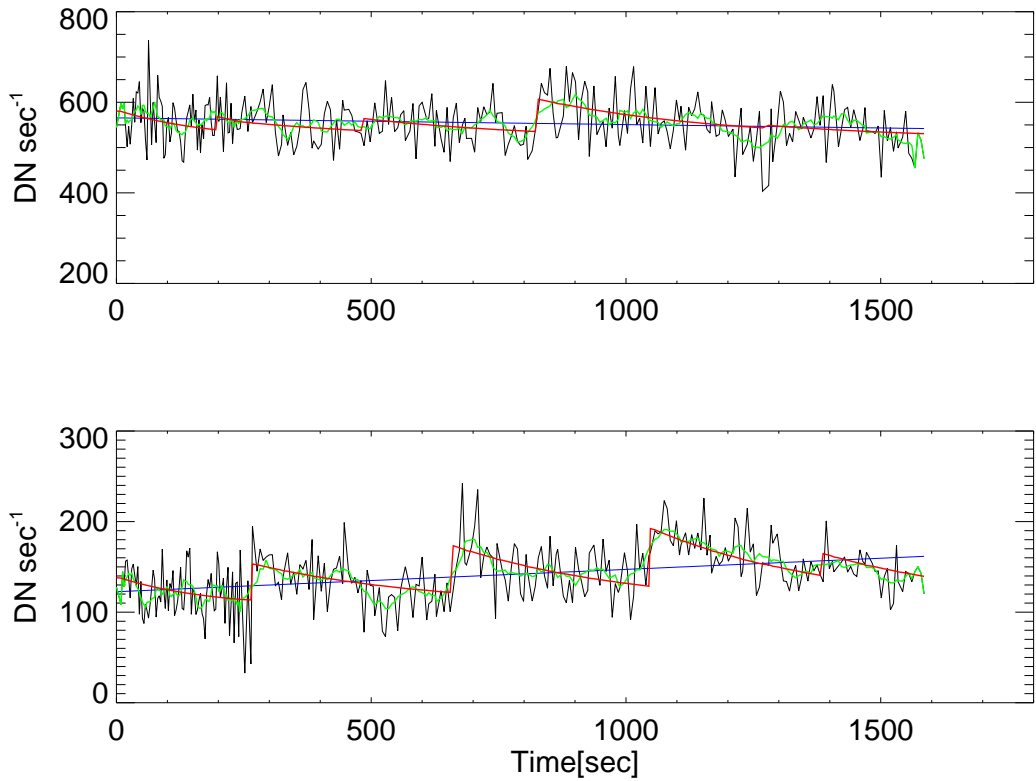}
\caption[]{Light curves of two pixels obtained from Monte Carlo simulations  adding trains of exponentials (red). The linear fits are marked (blue lines); 9-point ($\sim 1$ min) running averages are shown (green).}
\label{fig5}
\end{figure}

\begin{figure}
\epsscale{1.0} \plotone{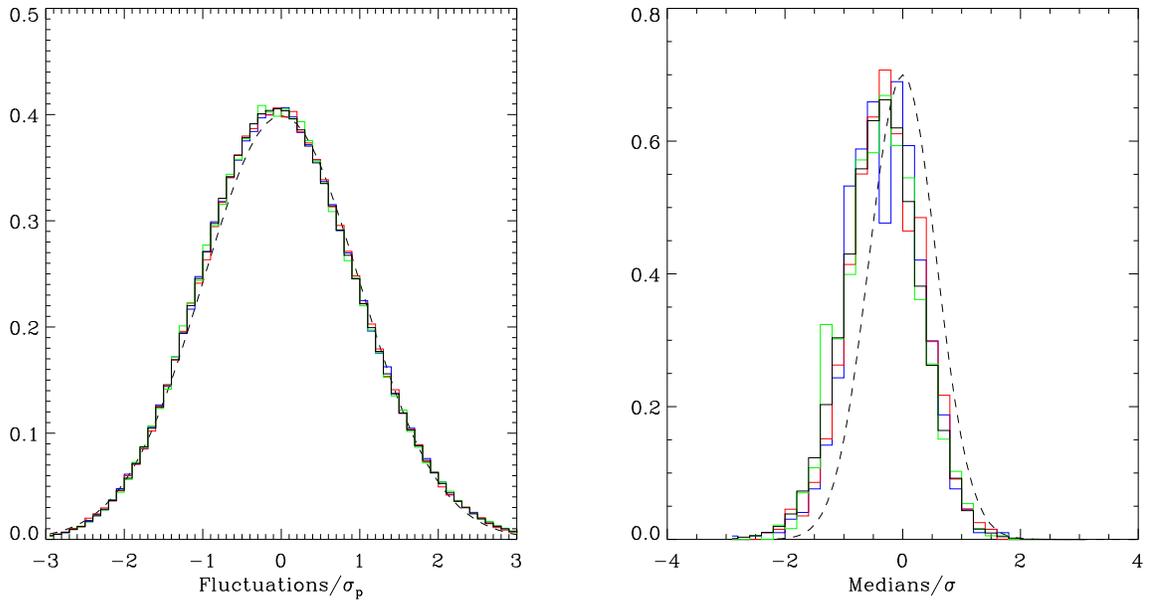} \caption[]{Identical to Figure
\ref{fig4} but obtained with the MonteCarlo simulation with: $A = 60$
DN/s, $\tau = 360$ s, and $dt = 360$ s.}
\label{fig6}
\end{figure}

The results of the simulations are listed in Tables \ref{tbl-2} and \ref{tbl-3}.
The median values from the simulations approach
those obtained from the data for all values of $\tau$, for $A = 60$
DN/s, and for time intervals of the order or larger than $\tau$
(Figures \ref{fig4} and \ref{fig6}). The best match with data
results is obtained with $A = 60$ DN/s, $\tau = 360\;s$ and $dt = 360\;s$.

It is worth commenting further on the distribution of median values
obtained from the individual pixels (Figures \ref{fig4} and
\ref{fig6}, right panels). As we have discussed, a negative
median is indicative of exponentially decreasing intensity and
cooling plasma (and also Poisson photon statistics to some degree).
However, a sizable fraction of the observed median values are
positive. Without the benefit of our simulations, we might conclude
that these pixels do not have cooling plasma.  The good agreement
between the observed (Fig.\ref{fig4}, right panel) and simulated
(Fig.\ref{fig6}, right panel) distributions, both in terms of
the centroid offset and the width, shows that the observations are
in fact consistent with all of the pixels having cooling plasma.
Positive median values occur when photon statistics mask the
relative weak signal of the exponentially decreasing intensity.

\subsection{Loop hydrodynamic modeling}
\label{sec:loopm}

\begin{figure}
\epsscale{.85}
\plotone{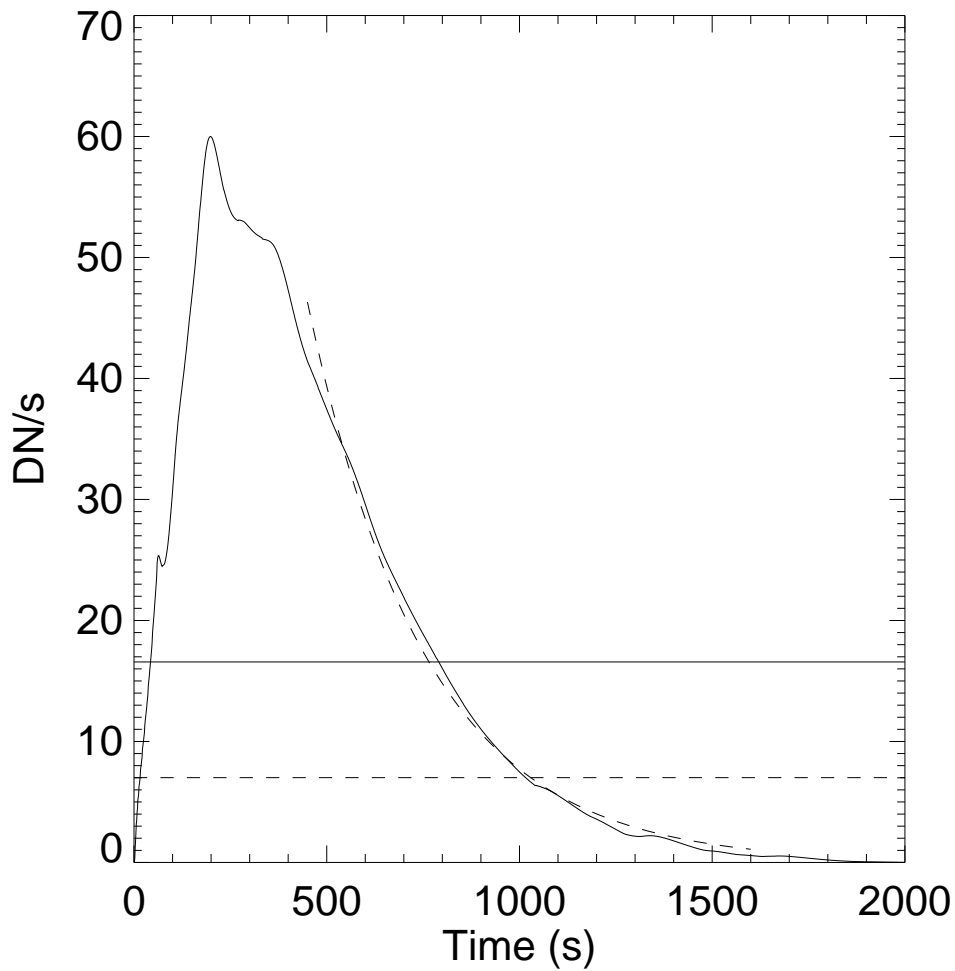}
\caption{Light curve in the XRT Al\_poly filterband obtained from a hydrodynamic simulation of the plasma confined in a loop strand ignited by a heat pulse (nanoflare). The heat pulse lasts 60 s and brings the strand to a maximum temperature $\log T \approx 7$. Most of the decay is well described by an exponential with an $e$-folding time $\tau \approx 300$ s (dashed line). Solid and dashed horizontal lines show
the mean and median intensity, respectively.}
\label{fig7}
\end{figure}

In a possible scenario, a coronal loop consists of many independent
strands, each ignited by a heat pulse that we call a nanoflare. The
evolution of the plasma confined in a single strand driven by a heat
pulse has been described in the past by means of time-dependent
hydrodynamic loop models
\citep{Nagai80,Peres82,Cheng83,Fisher85,MacNeice86}.  The light
curve in Figure~\ref{fig7} is synthesized in the Hinode/XRT Al\_poly
filterband from the results of a hydrodynamic model of a nanoflaring
strand \citep{Guarrasi010}. This hydrodynamic simulation has been
used successfully to explain totally different observational
results, which indicates that the parameters are realistic. The strand half-length is $3 \times 10^9$ cm. The heat
pulse of the single strand is a \emph{top-hat} function in time, the high state lasting $60\;s$, and in space it is uniformly distributed along the strand. Its intensity is 0.38 erg cm$^{-3}$ s$^{-1}$ and brings the strand to a maximum temperature $\log T \approx 7$. The total energy injected in the strand is therefore $\approx 1.4 \times 10^{11}$ erg cm$^{-2}$ to be multiplied by the strand cross-section area. The loop hydrodynamic simulations are
one-dimensional and in the synthesis of the loop emission the
cross-section area is a free parameter. We have chosen the
cross-section area so as to have an emission peak of 60 DN/s, a
realistic value suggested by the MonteCarlo simulations described
above. The light curve is characterized by a steep rise phase, a
short plateau and a much longer decay phase, which can be well
approximated by a decreasing exponential (Figure~\ref{fig7}). For this particular model strand (it depends on the strand half-length, see Eq.~\ref{eq:serio}), the best-fit $e$-folding time is $\sim 300$ s. We verified that the
median intensity (7.0 DN/s) is much smaller than (less than half of)
the mean intensity (16.6 DN/s). 

\section{Discussion}
\label{sec:disc} 

We find evidence that the light curves in each
pixel of an active region have systematic features: the distribution
of intensity fluctuations is asymmetric and the median value is less
than the mean. The effect is confirmed and even at higher level of significance 
when summed over larger and larger parts of the region, 
and therefore widespread and real.

We have also shown that part of the negative offset of the median
values is due to photon noise. We determine how much by performing
MonteCarlo simulations to generate synthetic light curves. Comparing the value
of the median for the entire region in Table~\ref{tbl-1} with the value of
the median for the simulations with Poisson noise only (\emph{null hypothesis}, $A = 0$, $threshold = 30$ in Table~\ref{tbl-2}) we see that the Poisson noise
accounts only for the $\sim 60$\% of the negative shift of the median.
The significance of the remainder is at the $5 \sigma$
level for the subregions and $25 \sigma$ level for the active
region!

We also perform simulations meant to represent cooling plasma by
randomly adding pieces of exponential decays onto the constant
background intensity. Photon noise is included as explained above.
The resulting light curves (see Figure \ref{fig5}) look similar to those in Figure
\ref{fig2}. The distributions of the intensity fluctuations agree
well with observations, with median values that have a similar
negative offset. As an aside, the parameters of the simulations lead
to realistic constraints about the loop substructuring (see the Appendix). 
We roughly estimate a possible strand diameter around $10^7$ cm, i.e. a fraction of arcsec, not far from the
resolution of the current instruments. Probably these are the most
significant nanoflare events, the high tail of a distribution. The
bulk of the events may occur with higher frequency and in finer
strands. 

We remark that our analysis is entirely independent of filter
calibration and highly model-independent. The data error is in
principle dependent on the emitted spectrum and therefore on the
plasma temperature and filter calibration, but we have estimated it
directly from the noise of the light curves. The model we use in
Monte Carlo simulations is very simple and has a minimal set of free
parameters.

Previous attempts to determine the nature of coronal heating outside
of isolated warm loops have been inconclusive (\citealp{Brooks_2009ApJ,Tripathi_2011ApJ}).
Our study provides strong evidence for widespread cooling plasma in active region AR
10923. This suggests heating that is impulsive and definitively
excludes steady heating, which in turn suggests that nanoflares play
a universal role in active regions. We favor nanoflares occurring
within the corona, but we do not exclude that our observations may
also be consistent with the impulsive injection of hot plasma from
below, as has recently been suggested \citep{De_Pontieu_2011Sci}.

\bigskip
\acknowledgements{We thank the anonymous referee for very useful suggestions. We also thank M. Caramazza and Y. Sakamoto for help in data analysis. Hinode is a Japanese mission developed and launched by ISAS/JAXA, collaborating with NAOJ as a domestic partner, NASA and STFC (UK) as international partners. Scientific operation of the Hinode mission is conducted by the Hinode science team organized at ISAS/JAXA. This team mainly consists of scientists from institutes in the partner countries. Support for the post-launch operation is provided by JAXA and NAOJ (Japan), STFC (U.K.), NASA, ESA, and NSC (Norway). F.R., S.Te. and M.M. acknowledge support from
Italian Ministero dell'Universit\`a e Ricerca and Agenzia Spaziale Italiana
(ASI), contracts I/015/07/0 and I/023/09/0. The work of J.A.K. was supported by the NASA Supporting Research and Technology and LWS Targeted Research and Technology programs.}

\appendix

\section{Loop substructuring}
\label{App.A}

We can make simple estimates of some characteristics implied by the
parameters constrained with MonteCarlo simulations. Let us assume
that the events that we resolve are able to heat an active region
loop, that an event observed in a pixel heats a whole loop strand,
that the intensity of each event is able to bring the loop to a
temperature of 10 MK, with an average temperature of 3 MK, and that
the loop has a total length of $2L = 5 \times 10^9$ cm.

From MonteCarlo simulations, we find that an appropriate average
event cadence interval is:
\[
dt \geq 360 ~~ s
\]

For an observation duration:
\[
\Delta t = 1600 ~~ s
\]

the number of events per pixel is:
\[
dn \approx \frac{\Delta t}{dt} \leq 4
\]

If we assume an average loop half-length ($10^9$ cm):
\[
L_9 \approx 2.5
\]

and a loop diameter (typically $10$\% of the loop length):
\[
D \approx 0.1 \times 2 L \approx 0.5 \times 10^9 {\rm cm} \approx 7 ~~{\rm pix}
\]

The number of events in the loop is:
\[
n \approx dn \times D \leq 30
\]

From loop scaling laws \citep{Rosner_1978ApJ} , we estimate the equilibrium pressure corresponding to the maximum temperature (MK) $T_{max,6}=10$, possibly due to a heat pulse:

\[
  p \approx 0.3 \frac{T_{0,6}^3}{L_9} \sim 100~  {\rm dyne ~ cm^{-2}}
\]

From this we roughly estimate the pulse heating rate per unit volume in units of $10^{-3} {\rm erg ~ cm^{-3} ~ s^{-1}}$ to bring a strand to a temperature of 10 MK:
\[
H_{-3}  = 3 p^{7/6} ~ L_9^{-5/6} \sim 300
\]

and the pulse energy flux over the whole loop:
\[
F = H \times 2 ~ L \sim 0.3 \times 5 \times 10^9 \sim 1.5 \times 10^9 ~~~ {\rm erg ~ cm^{-2} ~ s^{-1}}
\]

The energy released by the nanoflare in the loop is then:
\[
E_n = F~ n ~t_n~ dA_n \leq 1.5 \times 10^9 \times 30  t_n ~dA_n \approx 4 \times 10^{10} t_n~ dA_n
\]
where $t_n$ is the nanoflare duration and $dA_n$ is strand cross-section area.

Let's now consider the average loop conditions.
For a loop cross-section of:
\[
A = \pi ~R^2 \sim \pi 6 \times 10^{16} \sim 2 \times 10^{17} ~~ {\rm cm}^2
\]

and an average loop heating rate per unit volume for steady state ($T_{max,6}=3$):
\[
\langle H \rangle \sim 0.002~~ {\rm erg ~ cm^{-3} ~ s^{-1}}
\]

The loop total thermal energy in the observation can be estimated as:
\[
E_L \approx \langle H \rangle~2 ~L ~A ~\Delta t \approx 0.002 \times 2 \times 2.5 \times 10^9 \times 2 \times 10^{17} \times 1600 \approx 3 \times 10^{27} ~~~ {\rm erg}
\]

By equating $E_n \approx E_L$, we obtain:
\[
t_n \frac{dA_n}{A} \geq \frac{3 \times 10^{27}}{4 \times 10^{10} \times 2 \times 10^{17}} \approx 0.5
\]

So the product of the nanoflare duration (in s) and the fractional strand area is of the order of 1. For instance, if the nanoflare lasts 60 s we fill the loop with about 120 strands. The implication would be that the strand diameter is more than $10^7$ cm, but we warn that this is a crude estimate, based on scaling laws that hold only roughly out of equilibrium.

\begin{deluxetable}{crrr}
\tabletypesize{\scriptsize}
% \rotate
\tablecaption{Active Region Analysis results\label{tbl-1}}
\tablewidth{0pt}
\tablehead{
\colhead{Data} & \colhead{Threshold 30} & \colhead{Threshold 800} & \colhead{Threshold 1600} }
\startdata
Region         &-0.0258$\pm$0.0004 & -0.0160$\pm$0.0009 & -0.0136$\pm$0.0018 \\
Sub-reg 1      &-0.025$\pm$0.002  &  \nodata  &  \nodata \\
Sub-reg 2      &-0.026$\pm$0.002  &  \nodata  &  \nodata \\
Sub-reg 3      &-0.030$\pm$0.002  &  \nodata  &  \nodata \\
\enddata
\tablecomments{Table \ref{tbl-1} shows the values of the median averages, with errors, for the entire active region, and for the selected subregions (Fig.\ref{fig1}). The listed values for the entire active region are obtained analyzing only pixels with intensity over three different threshold values.}
\end{deluxetable}

\begin{deluxetable}{crrrrr}
\tabletypesize{\scriptsize}
% \rotate
\tablecaption{MonteCarlo Simulations results\label{tbl-2}}
\tablewidth{0pt}
\tablehead{
\colhead{$A^{\;1}$} & \colhead{$dt^{\;2}$} & \colhead{$\tau^{\;2}$} & \colhead{$Thr=30^{\;1}$} & \colhead{Thr=800} & \colhead{Thr=1600}}
\startdata
0  &  0  & 0 & -0.0164$\pm$0.0004  &  -0.0096$\pm$0.0009   &  -0.0096$\pm$0.0017 \\
% 10\% & 360 &   -0.0205$\pm$0.0005  &  -0.0181$\pm$0.0008   &  -0.0187$\pm$0.0017 \\
% 15\% & 360 &   -0.026$\pm$0.004    &  -0.0030$\pm$0.008    &  -0.0360$\pm$0.0016 \\
30$^{\;\;}$ & 360   & 360 & -0.0184$\pm$0.0004  &  -0.0105$\pm$0.0005   &  -0.0087$\pm$0.0017 \\
30$^{\;\;}$ & 540   & 360 & -0.0189$\pm$0.0004  &  -0.0099$\pm$0.0009   &  -0.0086$\pm$0.0018 \\
60$^{\;\;}$ & 360   & 180 & -0.0322$\pm$0.0004  &  -0.0136$\pm$0.0008   &  -0.0109$\pm$0.0017 \\
60$^{\;3}$  & 360   & 360 & -0.0253$\pm$0.0004  &  -0.0112$\pm$0.0009   &  -0.0070$\pm$0.0017 \\
60$^{\;\;}$ & 360   & 540 & -0.0228$\pm$0.0004  &  -0.0103$\pm$0.0009   &  -0.0063$\pm$0.0018 \\
60$^{\;\;}$ & 540   & 360 & -0.0283$\pm$0.0004  &  -0.0124$\pm$0.0008   &  -0.0087$\pm$0.0017 \\
\enddata
\tablecomments{Table \ref{tbl-2} shows the simulated values of averaged medians, with errors, for nanoflares heated active region. The cadence is Poisson-distributed around the average value, the amplitude is random-uniform between 0.5 and 1.5 the average value, A. A $= 0$ is the \emph{null-hypothesis} (no perturbation).}
\tablenotetext{1}{The amplitude of nanoflares and the threshold of intensity for the simulated pixels are in unit of $DN\;s^{-1}$.}
\tablenotetext{2}{The sampling spacing ($dt$) and the $e$-folding time ($\tau$) are in unit of seconds.}
\tablenotetext{3}{Simulation that best approaches the values measured in the observation.}

\end{deluxetable}

\begin{deluxetable}{crrrr}
\tabletypesize{\scriptsize}
% \rotate
\tablecaption{MonteCarlo Simulations results for Sub-regions\label{tbl-3}}
\tablewidth{0pt}
\tablehead{
\colhead{A} & \colhead{dt} & \colhead{Sub-reg 1} & \colhead{Sub-reg 2} & \colhead{Sub-reg 3}}
\startdata
0 &   0 &  -0.016$\pm$0.002  & -0.013$\pm$0.002   & -0.018$\pm$0.002 \\
30 & 360 &   -0.018$\pm$0.002  & -0.018$\pm$0.002 &  -0.018$\pm$0.002 \\
30 & 540 &   -0.021$\pm$0.002  & -0.017$\pm$0.002  & -0.020$\pm$0.002 \\
60 & 360 &  -0.021$\pm$0.002  &  -0.021$\pm$0.002 &  -0.024$\pm$0.002 \\
60 & 540 &  -0.024$\pm$0.002  &  -0.024$\pm$0.002  & -0.028$\pm$0.002\\
\enddata
\tablecomments{Table \ref{tbl-3} shows the simulated values of averaged medians, with errors, for selected sub-regions (Fig.\ref{fig1}) obtained from MonteCarlo simulations, with units as in Tab.\ref{tbl-2}.}
\end{deluxetable}

\bibliographystyle{apj}
% \bibliography{biblio}

\end{document}